\documentclass[a4paper]{article}

\usepackage{icrc2013}
\usepackage{color}
\usepackage{amsmath}
\title{Quark nuggets search using 2350 Kg gravitational waves
  aluminum bar detectors}

\shorttitle{Quark nuggets search using 2350 Kgr. gravitational waves
  aluminum bar detectors}

\authors{
P. Astone$^1$, M. Bassan$^2$,
E. Coccia$^2$, S. D'Antonio$^2$,
V. Fafone$^2$, 
G. Giordano$^3$, A. Marini$^3$, Y. Minenkov$^2$,
I. Modena$^2$, A. Moleti$^2$, G. V. Pallottino$^1$, G. Pizzella$^3$,
A. Rocchi$^2$, F. Ronga$^3$, M. Visco$^4$.
}

\afiliations{
$^1$ University of Rome "La Sapienza" and INFN, Rome1\\
$^2$  University of Rome "Tor Vergata" and INFN, Rome2\\
$^3$  Istituto Nazionale di Fisica Nucleare INFN, LNF Frascati\\
$^4$ IAPS-INAF and INFN, Rome2\\
}

\email{francesco.ronga@lnf.infn.it} 

\abstract{

The gravitational wave resonant detectors can be used as  detectors of quark nuggets, like nuclearites (nuclear matter  with a strange quark). This search has been carried out using  data from two  2350 Kg, 2 K cooled, aluminum bar detectors:
 NAUTILUS, located in Frascati (Italy), and EXPLORER, that was located in CERN Geneva (CH). Both antennas are equipped with cosmic ray shower detectors:  signals in the bar due to showers are  continuously detected and used to characterize the antenna performances. The bar excitation mechanism is based on the so called thermo-acoustic effect ,  studied  on dedicated experiments that use particle beams.  This mechanism predicts  that vibrations of bars   are induced by the heat  deposited in the bar from the particle.  The geometrical acceptance of the bar detectors is 19.5 $\rm m^2$ sr, that is smaller than that of other detectors used for similar searches. However, the detection mechanism is  completely different and is more straightforward than in other detectors.
We will show the results of ten years of data from NAUTILUS  (2003-2012) and  7 years from EXPLORER (2003-2009). 
The experimental  limits we obtain are of  interest because, for nuclearites of mass less than $10^{-4}$ grams, we find a flux  smaller than that one predicted considering  nuclearites  as   dark matter candidates.

}

\begin{document}
\maketitle

\section{Introduction}

Cosmic ray showers can excite mechanical vibrations in a metallic cylinder at its resonance frequencies and can provide an accidental background for experiments searching gravitational waves ($gw$): this possibility 
was suggested many years ago and a first search, ending with a null result, was carried out with room temperature Weber type resonant bar detectors \cite{Ezrow:1970yg}.
  
Later on, the cryogenic resonant $gw$ detector NAUTILUS~\cite{coccia} was equipped with a streamer tube extensive air  shower detector~\cite{coccia}
  and the interaction of cosmic ray with the antenna was studied in detail. This apparatus allowed the first detection  of cosmic ray
   signals in a $gw$ antenna, that took place in 1998, when NAUTILUS was operating at a temperature $T$~=~0.14~K~\cite{cosmico1},  i.e. below 
   the superconducting ($s$) transition critical temperature $T_c \simeq$~0.9~K.  During this run many events of very large amplitude were 
   detected\cite{cosmico2}. 

A detailed study of this effect is indeed useful to study the performance of $gw$ bar detectors for exotic particles~\cite{Astone:1992zf} like nuclearites, and  to understand the noise due to cosmic rays in interferometric $gw$ detectors~\cite{Yamamoto:2008fs}. In this paper we will report on an update of the results obtained on nuclearites in ref.~\cite{Astone:1992zf}  with an increase on the exposure  of about a factor 30. 

Recently the possibility to have 
compact ultradense quark nuggets objects   has been stressed  again, see for example reference  \cite{Labun:2011wn}. 
Probably the negative dark matter searches in LHC and in direct and indirect experiments  pushed in this direction.
Nuclearites are an example of compact objects that could be constituent of the dark matter; the results described in this paper are therefore of more general interest. More informations on the nuclearites detection are in another paper at this conference \cite{jemeuso}.

\section{The  NAUTILUS and EXPLORER $gw$ bar detectors}

The $gw$  detector NAUTILUS\cite{nautilus} is located in Frascati (Italy) National Laboratories of INFN,  at about 200 meters above sea level.
NAUTILUS started  operations around 1998. The current  run started in 2003.




The  detector EXPLORER   \cite{explorer} 
was located in CERN (Geneva-CH) at about 430 meters above sea level. The EXPLORER run ended in June 2010. 



Both detectors use the same principles of operation.   EXPLORER and NAUTILUS consist of a large aluminum alloy cylinder (3 m long, 0.6 m diameter) suspended in vacuum by a cable around its central section and cooled to about 2 K by means of a superfluid helium bath.
The ($gw$)  excites the odd longitudinal modes of the cylindrical bar, which is cooled to cryogenic temperatures to reduce the thermal noise
 and is isolated from seismic and acoustic disturbances.
  To record the vibrations of the bar first longitudinal mode, an auxiliary mechanical resonator tuned to the same frequency is bolted on one bar end face. This resonator is part of a capacitive electro-mechanical transducer that produces an electrical a.c. current that is proportional to
  the displacement between the secondary resonator and the bar end face. Such current is then amplified by means of a dcSQUID 
  superconductive device. NAUTILUS is also equipped with a dilution refrigerator that enables operations at 0.1 K, further reducing the thermal noise. In the period considered, however, the refrigerator was not operational, in order to maximize the detector duty cycle. 
  
   Both detectors are equipped with cosmic ray telescopes, to veto excitations due to large showers. The two telescopes rely on different technologies (scintillators for Explorer, streamer tubes for NAUTILUS) but both provide a monitor of comparable effectiveness and a continuous check of the antenna sensitivity.
  
  The output of the SQUID amplifier is conditioned by band pass filtering and by an anti-aliasing low-pass filter, then sampled at 5 kHz and stored on disk. Sampling is triggered by a GPS disciplined rubidium oscillator, also providing the time stamp for the acquired data.
The data are processed off-line, applying adaptive, frequency domain filters. We first ÓwhitenÓ the data, i.e. remove the effect of the detector transfer function.  A filter matched to delta (or very short) excitations is then applied to this stream. The filter is designed and optimized for delta-like signals, but it works equally well  for a wider class of short bursts, like e.g. damped sinusoids with decay time less than 5 msec.  The noise characteristics estimate is updated averaging the output over 10 minutes periods. Traditionally the noise is expressed as energy in Kelvin units. The typical noise of  data considered in this paper is between 1 and 5 mK.
  
  At present, while the large interferometers VIRGO and LIGO are undergoing massive overhauls to upgrade their sensitivity, there are  still two resonant detectors, NAUTILUS and a similar detector AURIGA, that continue to operate in Óastro-watchÓ mode, i.e. as sentinels recording data that could be analyzed in conjunction with a significant astrophysical trigger, such as the explosion of a nearby supernova, or any astronomical event thought to be a possible source of $gw$.

\section{The thermo-acoustic model }
 
The interaction of energetic charged particles with a normal mode of an extended elastic cylinder has been extensively studied over the years, both on the theoretical and on the experimental aspecta.

The first experiments aiming to detect mechanical oscillations in metallic targets due to impinging elementary particles were carried out by Beron and Hofstander as early as in 1969 \cite{beron1}.
A few years later, Strini et al. \cite{grassi} carried out an experiment with a small metallic cylinder and measured the cylinder oscillations.  The authors compared the data against the thermo acoustic model in which the longitudinal vibrations are originated from the local thermal expansion caused by the warming up due to the energy lost by the particles crossing the material. In particular, the vibration amplitude is directly proportional to the ratio of two thermophysical parameters of the material,  namely the thermal expansion coefficient and the specific heat at constant volume. The ratio of these two quantities appears in the definition of  the Gr\"{u}neisen parameter $\gamma$. It turns out that while the two thermophysical parameters vary with temperature, $\gamma$ practically does not, provided the temperature is above the material superconducting $(s)$ state  critical temperature.  

\begin{figure}
  \begin{center}
  \includegraphics[height=60mm,width=80mm]{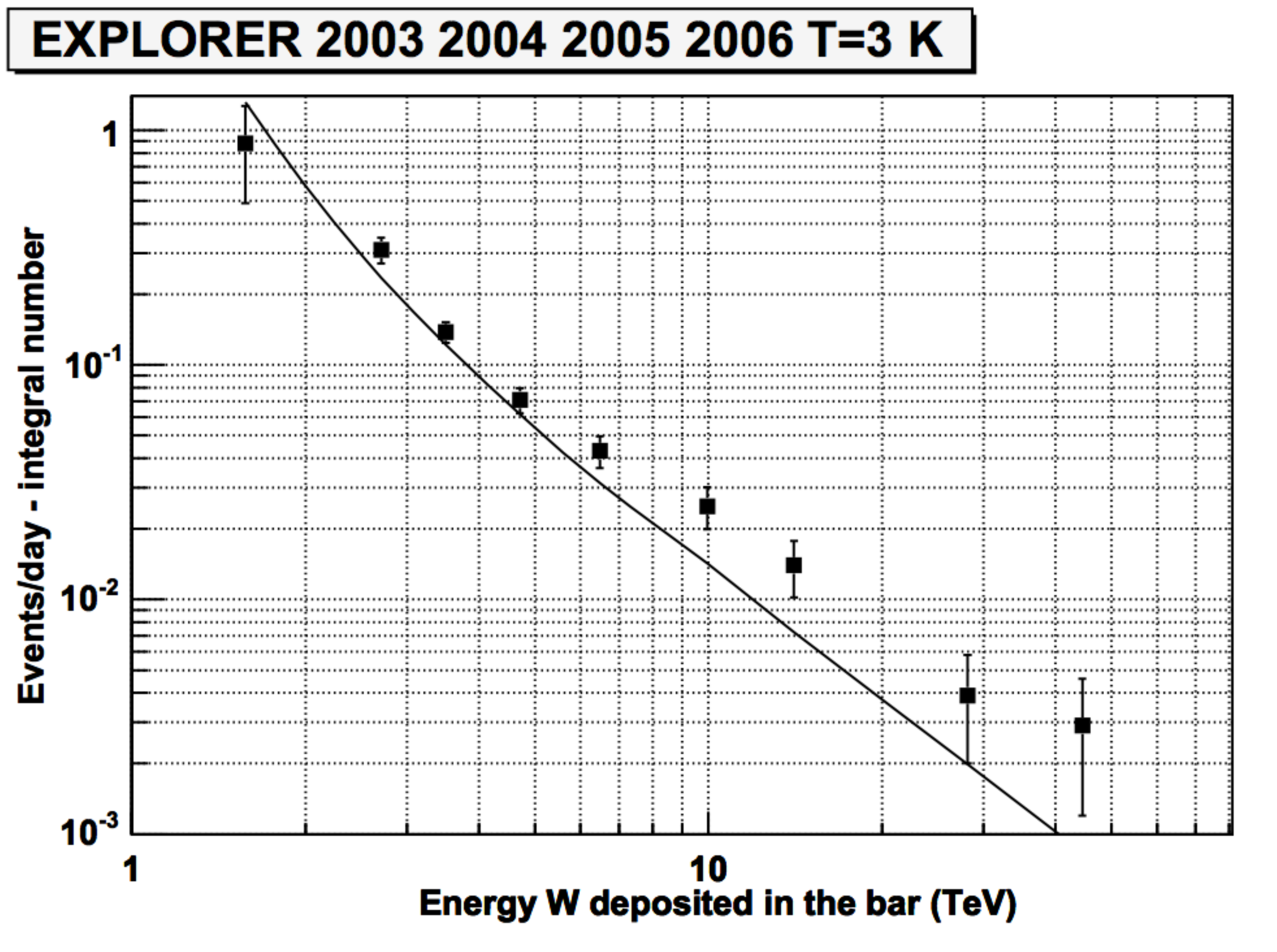}
  \end{center}
  \vspace{-0.5cm}
  \caption{Integral distribution of extensive air showers in Explorer, the line shows the prediction based on the thermo acoustic model \cite{Astone:2008xa}. The largest event has 360 TeV. ( 670 K in Kelvin units)}
  \label{fig:shower}
\end{figure}

Detailed calculations,  successively refined by several authors \cite{allega,deru,liu} agree in 
predicting, for the excitation energy  $E$ of the fundamental vibrational mode of an aluminum cylindrical bar, the following equation: 

\begin{multline}
 E=\frac{4}{9\pi}\frac{\gamma^2}{\rho L v^2}(\frac{dW}{dx})^2 \times \\
 \times [sin(\frac{\pi z_o}{L})\frac{sin[(\pi l_ocos(\theta_o)/2L]}{\pi Rcos(\theta_o)/L}]^2
 \label{eliub}
\end{multline}

where  $L$ is the bar length, $R$ the bar radius, $l_o$ the length of the particle track inside the bar, $z_o$ the distance of the track mid point  from one end of the bar, $\theta_o$ the angle between the particle track and the axis of the bar, $\frac{dW}{dx}$ the energy loss of the particle in the bar, $\rho$ the density, $v$ the  longitudinal  sound velocity in the material.
This relation  is valid for the  normal-conducting $(n)$ state material and some authors (see ref.  \cite{allega,deru}) have extended the model to a super-conducting  ($s$) resonator, according to a scenario in which the vibration amplitude is due to two pressure sources, one due to $s-n$ transitions in small regions centered around the interacting particle tracks and the other due to thermal effects in these regions now in the $n$ state.

 It is important to note, at this point, that  a $gw$ bar antenna, used as particle detector, has characteristics very different from the usual particle detectors which are sensitive to ionization losses:  indeed an acoustic resonator can be seen as a zero threshold calorimeter, sensitive to a vast range of energy loss processes. The usual $gw$ software filter works well up to a time scale of the order of the order of 5 msec, corresponding to a $\beta=4 \times 10^{-6}$ for a 60 cm particle track. So the antenna is sensitive  to very slow tracks: this is another very important difference with to the usual particle detectors.

\begin{figure}
  \begin{center}
  \includegraphics[height=50mm,width=80mm]{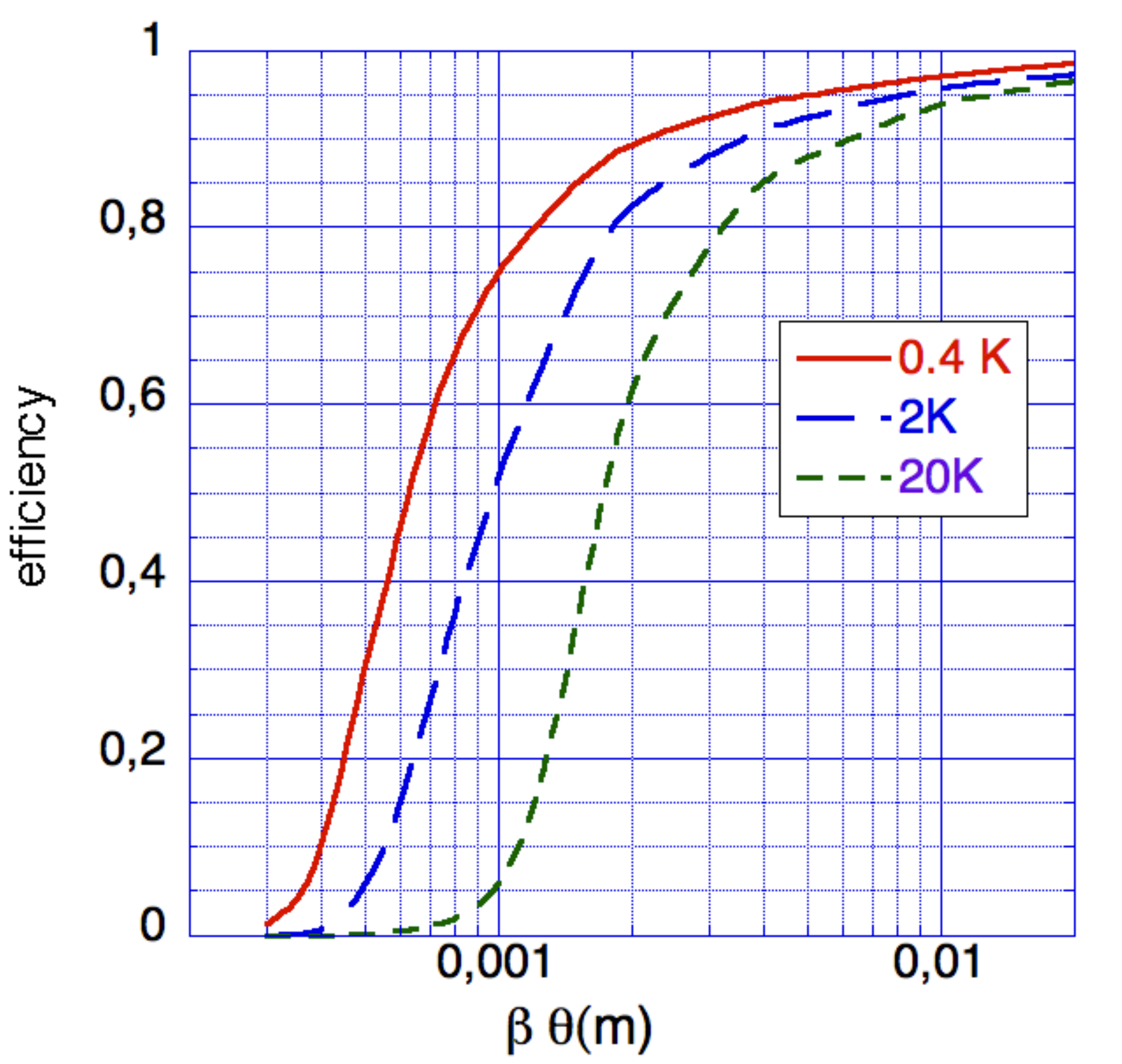}
  \end{center}
  \vspace{-0.5cm}
  \caption{Efficiency respect to the geometrical acceptance as functions of $\beta \theta(M)$ for 3 different thresholds of energy detection (in Kelvin).}
  \label{fig:effic}
\end{figure}

As anticipated in the introduction, the first detection of signals in a  detector output due to cosmic ray events, took place in  1998 with NAUTILUS  at $T=0.14$ K~\cite{cosmico1},  i.e. below the $s$ transition
 temperature $T_c \simeq0.9 K$.
 and many  events of unexpectedly large amplitude were detected. This result suggested an anomaly either in the model or in the cosmic ray interactions\cite{cosmico2}. However the observation was not confirmed in the 2001 run with NAUTILUS at $T=1.5$ K~\cite{cosmico3} and therefore we made the hypothesis that the unexpected behavior was due to the superconducting state of the material. 
 An extended paper on this argument has been published \cite{Astone:2008xa} and the
  results of a dedicated experiment on an electron particle measured an enhancement of a factor 24 (in  energy)  of the  signals at  a bar temperature $T=0.14 K$  \cite{Nim2011}.
 Now we have a good agreement both in rate and amplitude of the extensive ray shower detected in NAUTILUS and EXPLORER as shown in Fig.\ref{fig:shower}, 
 with the expectation based on  cosmic ray physics and the thermo acoustic model. Therefore  we are confident that we have a full understanding
  of the $gw$ bar detectors used as particle detector.

\section{Nuclearite search in NAUTILUS and EXPLORER}
According to \cite{deru1,Witten84} nuclearites are considered to be large
strange quark nuggets, with overall neutrality ensured by an
electron cloud which surrounds the nuclearite core, forming a sort
of atom. Nuclearites with galactic velocities are protected by their
surrounding electrons against direct interactions with the atoms
they might hit.

\begin{figure}
  \begin{center}
  \includegraphics[height=50mm,width=80mm]{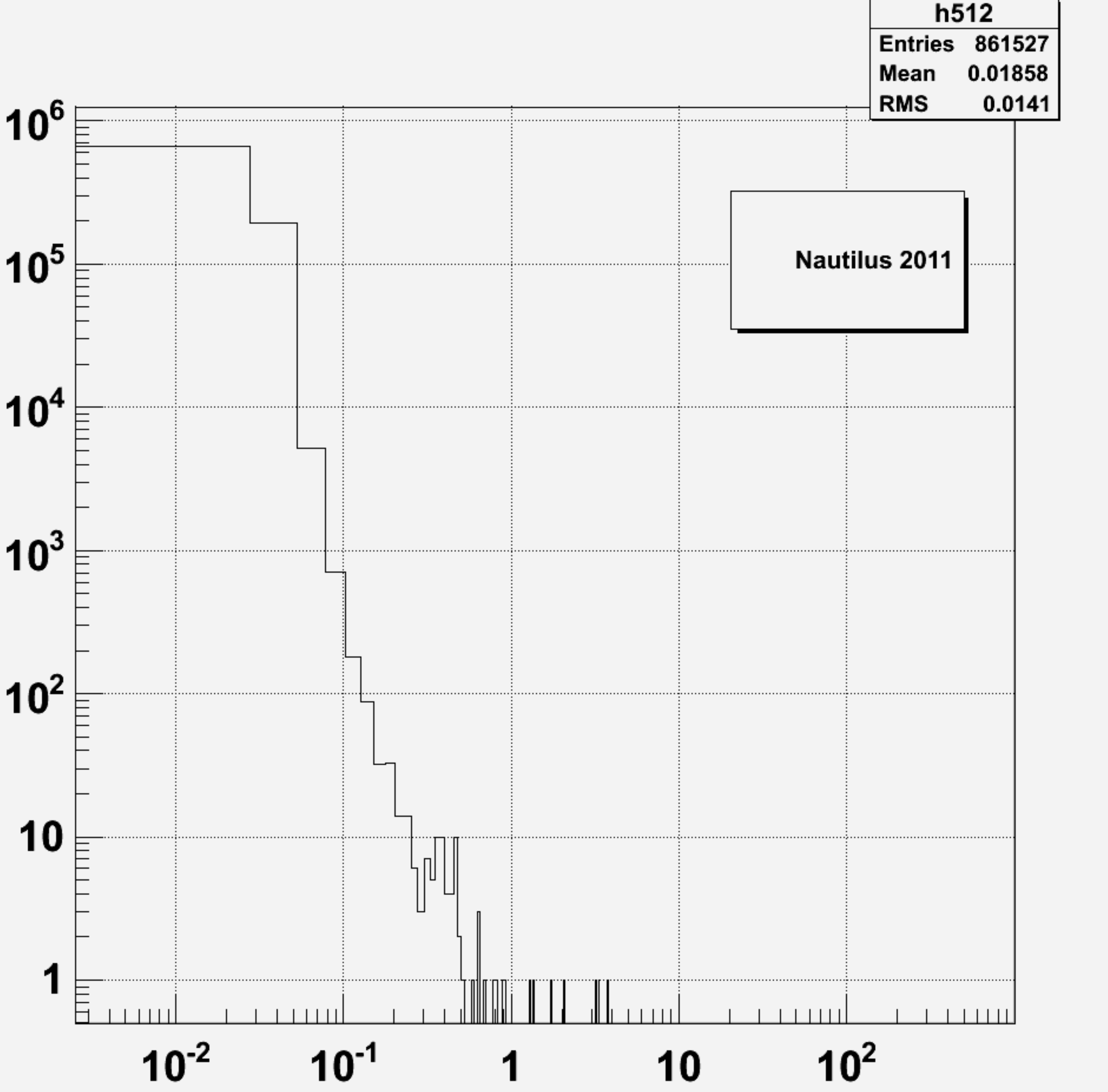}
  \end{center}
  \vspace{-0.5cm}
  \caption{Energy distribution (in Kelvin units), of the "best" run year 2011 of NAUTILUS. The energy of the biggest event is 3.7 K. This event is a cosmic ray (Extensive Air Shower).) }
  \label{fig:flux}
\end{figure}

As a consequence, the principal energy-loss mechanism for a
nuclearite passing through matter is atomic collision. For a massive
nuclearite the energy-loss rate is\cite{deru1}:

\begin{equation}
\label{eq:rel}
\frac{dE}{dx}= -A \rho v^{2}
\end{equation}
where $\rho$ is the density of the traversed medium, $v$ the
nuclearite velocity and $A$ is its effective cross-sectional area.
The effective area can be obtained by the nuclearite density
$\rho_{N}$. For a small nuclearite of mass less than $1.5$ $ng$, the
cross-section area $A$ is controlled by its electronic atmosphere
which is never smaller than $10^{-8}$ cm:

\begin{equation}
\label{eq:sigma}
A =\left\{ \begin{array}{ll}
\pi \cdot {10}^{-16} \,\textrm{cm$^{2}$}    &  for ~ M < 1.5 \, ng
  \\
\pi {\left(\displaystyle{\frac{3 M}{4 \pi {\rho}_{N}} }\right)}^{2/3} & for ~ M > 1.5 \, ng
\end{array} \right.
\end{equation}
where \mbox{$\rho_{N}= 3.5\cdot10^{14}\,g/cm^{3}$}  and $M$ is the nuclearite mass.

According to Eq. \ref{eq:rel}, nuclearites having galactic velocity
and mass heavier than $10^{-14}$ $g$ penetrate the atmosphere, while
those heavier than 0.1 $g$ pass freely though an Earth diameter. Equation  \ref{eq:rel} breaks down in a solid at velocity
smaller than the sound velocity in the medium; in aluminum this correspond at $\beta=2*10^{-5}$; for subsonic velocity
the energy loss becomes a constant and the nuclearite is rather quickly brought to  rest.

 Inserting Eq. \ref{eq:rel} in Eq. \ref{eliub} we obtain
the energy in the fundamental  mode of a cylindrical bar, that is the energy detected in gravitational $gw$ bar detectors.
Using the thermo acoustic parameters at T=2K   \cite{Nim2011} we have
 for a vertical nuclearite  of mass M and velocity $c\beta$ in the middle of the NAUTILUS (or EXPLORER) bar:
\begin{equation}
\label{eq:nume}
\Delta E[Kelvin]= 10.7 (\frac{ \beta \theta(M)}{10^{-3}})^4  
\end{equation}
where $ \Delta E$ is the energy variation of of bar fundamental mode measured in Kelvin  and
$\theta(M)=(M/1.5~ngr.)^{1/3}$ if M$>1.5$ ngr. otherwise $\theta(M)=1$

The maximum geometrical acceptance for a nuclearite isotropic distribution is given by $2\pi S_{tot}=19.54~m^2 sr^{-1}$, where $S_{tot}$ is the bar surface. The effect of the track path length and angle has been computed for an isotropic distribution by a Montecarlo.
The results  as function of $\beta \theta(M)$ and for different  $\Delta E$ thresholds are in Fig.\ref{fig:effic}.

\begin{figure}
  \begin{center}
  \includegraphics[height=70mm,width=80mm]{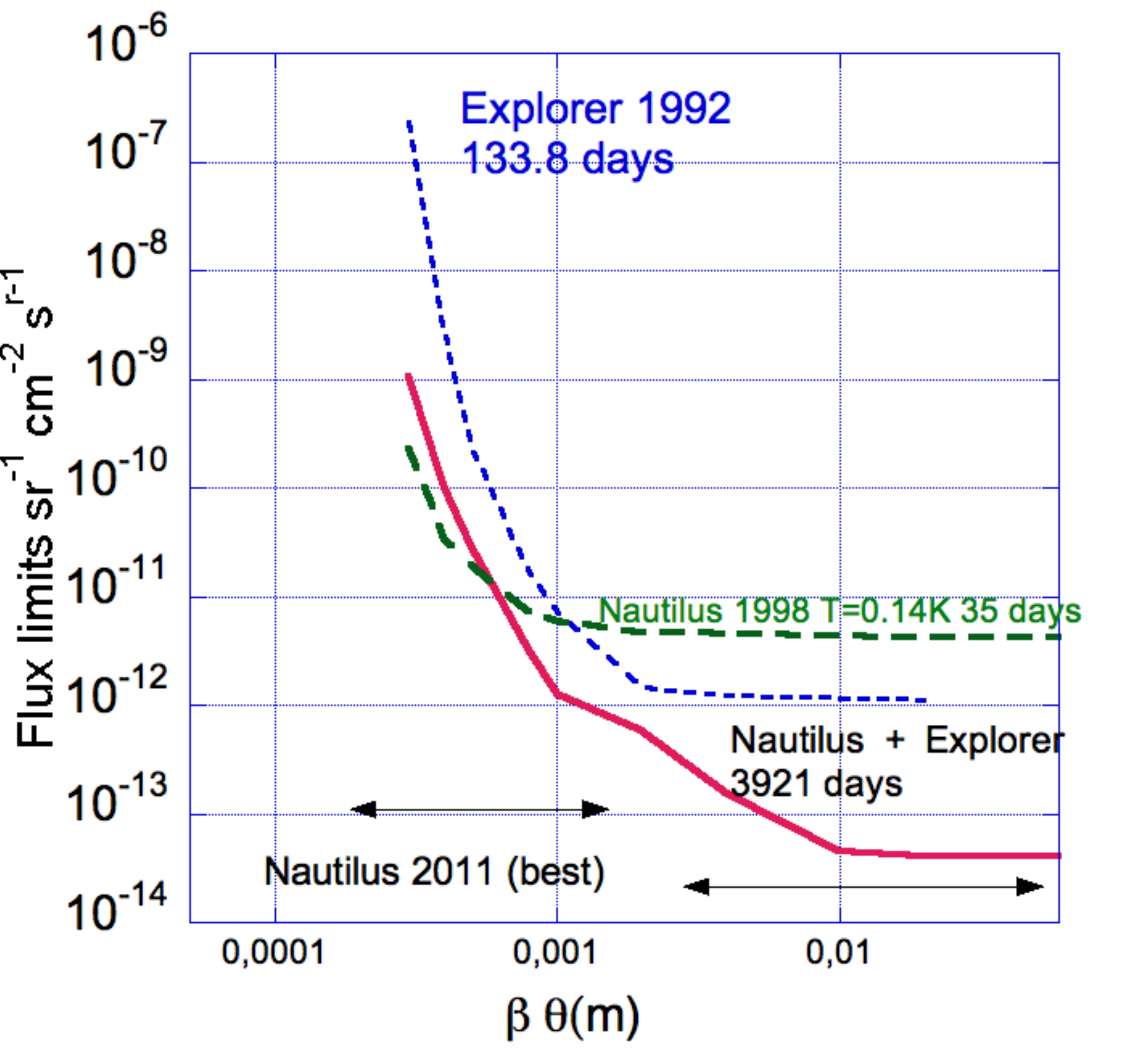}
  \end{center}
  \vspace{-0.5cm}
  \caption{90\% C.L. isotropic  flux upper limits compared to  previous results (Explorer) \cite{Astone:1992zf}. For nuclearites that can not penetrate the Earth there is a factor 2 in the flux limit. Limits  from a short run with NAUTILUS at $T=0.14$K may be interesting because of the different detection mechanism  in the superconducting state. }
  \label{fig:flux}
\end{figure}

For this search we have used the standard  filter matched to delta. In order to reduce noise we have applied several "standard" cut to the data.
The most important are: the gain of the electronic chain, the SQUID  locking working point, the noise outside the useful bandwidth of the detector,
the seismic monitors. In addition we have put  a cut on the run length, requiring at least 10h and a cut on the noise of the filtered data.
    The noise cut    is  $T_{eff}<5~mk$ for EXPLORER and  $T_{eff}<2.5~mK$ for NAUTILUS. The total live-time with those cuts is 2089.9 days
 for NAUTILUS and 1831.4 days for EXPLORER.

The nuclearite flux upper limits have been computed starting from the energy distributions.  Since the antenna noise has not-gaussian tails and changes with the run conditions we have identified the data stretches with the smallest noise, dividing  data in years. The best data-set is the Nautilus run in 2011. Therefore the 90\%C.L.  flux limit has been computed using the  Nautilus 2011 run, live-time=305.9 days  for small amplitude signals $\beta\theta(m)\le0.002$.  Otherwise for  $\beta\theta(m)> 0.002$ we have used the full data set with a total live-time=3921.3 days. The results are in Fig.~\ref{fig:flux}.
This figure shows also the limits  from a short run with NAUTILUS at $T=0.14$K , with live-time= 35.1 days,  this result may be interesting because of the different detection mechanism  in the superconducting state.
For   $\beta\theta(m)>0.01$ where the background is negligible the flux upper limit is dominated only by the live-time. Note that in this search events in coincidence with the cosmic ray detector are not removed. This is because fast nuclearites could produce light in the Explorer scintillators (due to black body emission\cite{deru1}) and could be confused with a cosmic ray event.

\begin{figure}
  \begin{center}
  \includegraphics[height=70mm,width=80mm]{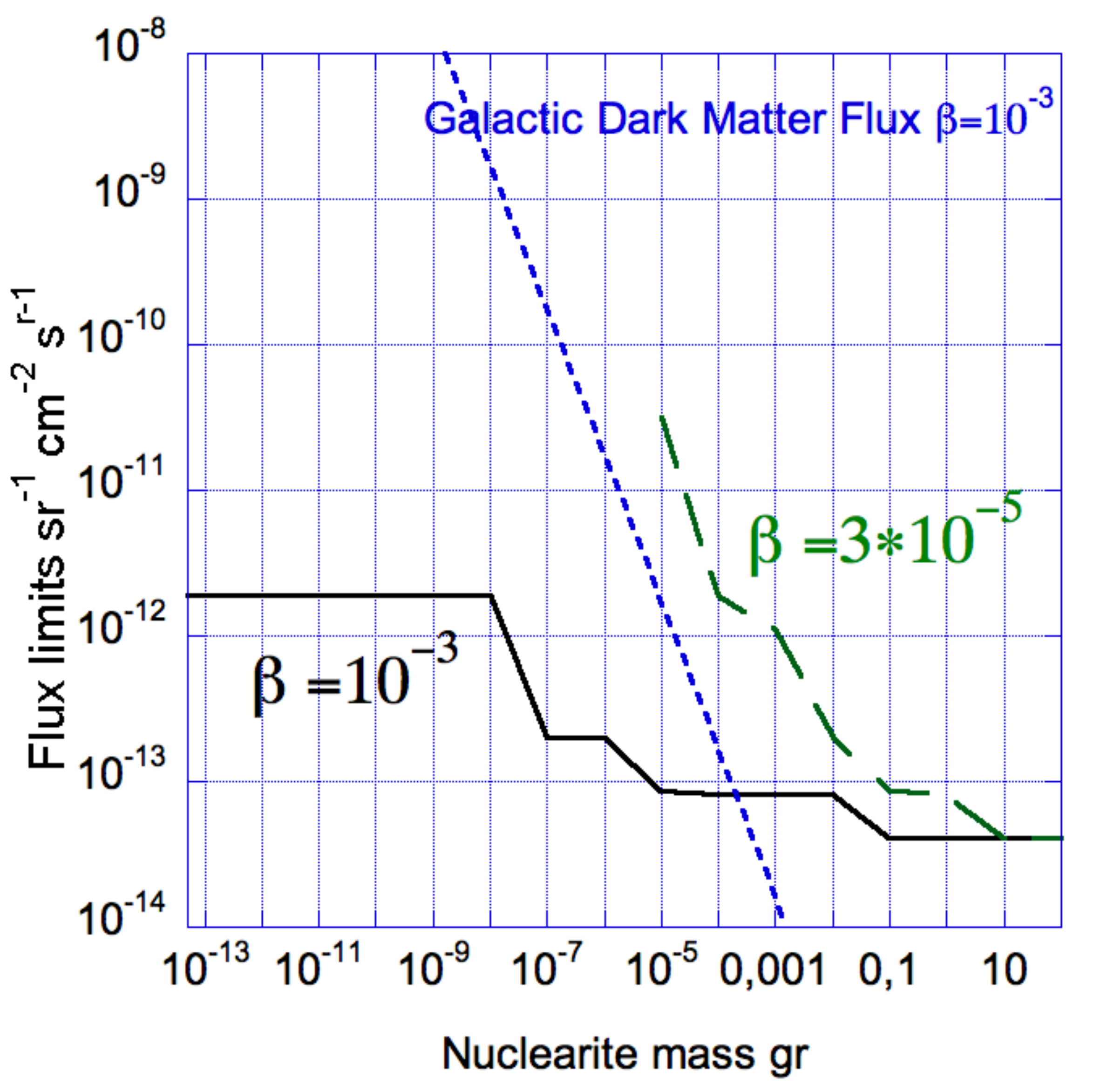}
  \end{center}
  \vspace{-0.5cm}
  \caption{Flux upper limits for $\beta=10^{-3}$ and  $\beta=3\cdot 10^{-5}$  (Earth escape velocity) vs  mass. }
  \label{fig:fluxMass}
\end{figure}

Finally  Figure~\ref{fig:fluxMass}  shows the upper limits vs f the nuclearite mass and for $\beta=10^{-3}$ typical of the galactic dark matter.
For mass$\le10^{-4}$ and  mass $\ge5\cdot 10^{-14}$ gr.  (threshold due to the atmosphere) this limit is significantly smaller than the flux of galactic dark matter. Earth is transparent for nuclearite of mass $\ge0.1$ gr., this produces a factor 2 reduction in the flux limit. Figure~\ref{fig:fluxMass} also shows the limits  for $\beta=3\cdot10^{-5}$,  the Earth escape velocity.
Those limits are derived from  Fig~\ref{fig:flux}, computing the appropriate $\beta\theta(M)$.

Other experiments  above sea level using track etch detectors have obtained lower limits.  The SLIM limit\cite{SLIM} for $ \beta=10^{-3}$ is   $1.3 \cdot 10^{-15}  cm^{-2} s^{-1} sr^{-1}$. and the OHYA\cite{Orito:1990ny} limit is $3.2 \cdot 10^{-16}  cm^{-2} s^{-1} sr^{-1}$.
There is no quantitative theory describing  the track etch mechanism.
Track etch detectors have been calibrated with slow charged ions, assuming energy lost  by Coulomb elastic collisions.  In principle this process is different from the energy loss of Eq.\ref{eq:rel}.


\section{Conclusions}
 
The energy loss predicted for compact ultradense quark nuggets  DM particles varies in different models, but  the main energy loss
mechanism is similar to one of nuclearites given by  Eq. \ref{eq:rel}.  With $gw$  bar detectors we directly  measure in a calorimetric way this energy. This 
technique has been verified on a particle beam and is used to continuously  monitor  the antenna performance using the extensive air showers in the cosmic rays.
Therefore our results on nuclearites are more general and could be applied to other compact objects\cite{Labun:2011wn}.

Our data analysis is still in progress: in particular we are looking for additional signatures to separate genuine delta like events from the noise and we are trying different optimizations of the  the data quality cuts.  The efficiencies of those cuts can be measured using cosmic rays.

\end{document}